# Phonon-limited Mobility of 2D Semiconductors: Quadrupole Scattering and Free-carrier Screening

Chenmu Zhang, Yuanyue Liu*

*Texas Materials Institute and Department of Mechanical Engineering*

*The University of Texas at Austin, Austin, Texas 78712, USA*

*Yuanyue.liu@austin.utexas.edu*

**Abstract:**

Two-dimensional (2D) semiconductors have demonstrated great potential for next-generation electronics and optoelectronics. An important property for these applications is the phonon-limited charge carrier mobility. The common approach to calculate the mobility from first principles relies on the interpolation of the electron-phonon coupling (EPC) matrix. However, it neglects the scattering by the dynamical quadrupoles generated by phonons, limiting its accuracy. Here we present a first-principles method to incorporate the quadrupole scattering, which results in a much better interpolation quality and thus a more accurate mobility as exemplified by monolayer $MoS_2$ and InSe. This method also allows for a natural incorporation of the effects of the free carriers, enabling us to efficiently compute the screened EPC and thus the mobility for doped semiconductors. Particularly, we find that the electron mobility of InSe is more sensitive to the carrier concentration than that of $MoS_2$ due to the stronger long-range scattering in intrinsic InSe. With increasing electron concentration, the InSe mobility can reach ~4 times of the intrinsic value, then decrease owing to the involvement of heavier electronic states. Our work provides accurate and efficient methods to calculate the phonon-limited mobility in the intrinsic and doped 2D materials, and improves the fundamental understanding of their transport mechanism.

**Introduction:**

Two-dimensional (2D) crystalline semiconductors are semiconducting materials with a thickness of only one or few atomic layer(s). The common 2D semiconductors include transition metal dichalcogenides such as $MoS_2$ [1,2], black phosphorus (BP) [3], and indium selenide (InSe) [4]. The extreme thinness of 2D semiconductors introduces many useful properties, such as efficient control of the properties by electrostatic gating [5,6], optical transparency [7], and mechanical flexibility [8,9]. These properties make the 2D semiconductors promising candidates for various electronic applications. However, the current 2D semiconductors suffer from relatively poor carrier transport properties at room temperature. For example, the mobilities of charge carriers (electrons and holes) of current 2D semiconductors with monolayer thickness are generally low at room temperature. They are typically less than 300 $cm^2V^{-1}s^{-1}$ [2,4,10,11]. Conversely, the electron mobility of silicon is ~ 1400 $cm^2V^{-1}s^{-1}$ [12,13], germanium is ~ 3900 $cm^2V^{-1}s^{-1}$ [14,15], and indium arsenide is ~ 30,000 $cm^2V^{-1}s^{-1}$ [16]. In order to better understand the mobility and design/discover materials with higher mobility, it is important to have an effective method to calculate the mobility from first principles.

Phonons are important scattering source at room temperature. There are several approaches to calculate the phonon-limited mobility. Models with simplified electron-phonon coupling (EPC) (e.g. deformation potential theory [17–22]) cost minimal computational resources, however, they can be very inaccurate and can significantly overestimate the mobility [23]. Another approach is to evaluate all the EPC matrix elements using density functional perturbation theory (DFPT) [24,25], which gives the most accurate mobility [26]. But such approach is computationally very expensive because one needs to calculate a large amount of EPC matrix elements (with a rather dense grid to sample the Brillouin zone) in order to get the converged mobility. Such high computational cost limits its application. A more common approach is to (1) first use DFPT to calculate the phonon-induced perturbation potentials and the corresponding EPC matrix elements on a sparse grid; (2) formulate an explicit expression for the long-range (LR) perturbation potentials and their contributions to the matrix elements, and subtract them from the calculated total matrix elements to get the short-range (SR) components; (3) interpolate the SR parts to a dense grid based on maximally localized Wannier functions [27], use the derived expression to get the corresponding LR parts, and sum them up to recover the total matrix elements [28]. This "interpolation" approach can afford a rather dense grid and a large amount of EPC matrix elements, and thus is more commonly used.

The accuracy of the interpolation approach, particularly for the EPC by the long-wavelength phonons, heavily relies on the treatment of the LR perturbation potential. The leading term in the LR perturbation potential is called Fröhlich or dipole potential and the next-to-leading order term is usually called quadrupolar potential. The EPC by the dipolar [28] and quadrupolar potentials [29–32] have been studied for 3D crystals; while for 2D crystals, only the dipolar scattering is considered [33–35], and the impact of quadrupole scattering remains elusive. In this work, we study the quadrupole scattering in 2D crystal semiconductors. We first present a first-principles method to calculate the LR EPC matrix element that includes the quadrupolar potential. Using the examples of $MoS_2$ and InSe, we show that the incorporation of the quadrupolar potential significantly improves the interpolation of the EPC matrix elements for long-wavelength phonons. Consequently, the calculated mobility is more accurate. We also provide an efficient approach to compute the screened EPC and thus the mobility for doped semiconductors. We find that the InSe electron mobility is more sensitive to the carrier concentration than that of $MoS_2$ as it has stronger long-range scattering in intrinsic InSe. When increasing electron concentration, the InSe mobility first increase due to the screening and then then decrease owing to the involvement of heavier electronic states. Significantly, it can reach 3.8 times of the intrinsic mobility (117 to 444 $cm^2V^{-1}s^{-1}$).

**Methods:**

The carrier mobility $\mu$ at low electric field can be obtained from the Boltzmann transport theory under momentum relaxation time approximation:

$$\mu_{\alpha\beta} = \frac{q}{n_c \Omega} \sum_n \int \frac{d\mathbf{k}}{\Omega_{BZ}} \frac{\partial f_{n\mathbf{k}}}{\partial E_{n\mathbf{k}}} \tau_{n\mathbf{k}} v_{n\mathbf{k},\alpha} v_{n\mathbf{k},\beta}, \tag{1}$$

where $\alpha$ and $\beta$ are the direction indices, $q$ is the charge of carrier, $\Omega$ ($\Omega_{BZ}$) is the area of unit cell (Brillouin zone); $\tau_{n\mathbf{k}}$ is the momentum relaxation time for the electronic state with band index $n$ and wavevector $\mathbf{k}$, $v_{n\mathbf{k}}$ is its group velocity, and $E_{n\mathbf{k}}$ is its energy; $f$ is the Fermi distribution function, and $n_c$ is the carrier density which is related with $f$ and the electronic band structure through:

$$n_e = \sum_n \int \frac{d\mathbf{k}}{\Omega_{BZ}} f_{n\mathbf{k}}, \quad n_h = \sum_n \int \frac{d\mathbf{k}}{\Omega_{BZ}} (1 - f_{n\mathbf{k}}), \tag{2}$$

where $n_e$ and $n_h$ are the concentrations for electrons and holes respectively. For 2D crystal with hexagonal lattice, the mobility tensor is a diagonal matrix with identical diagonal matrix elements. In this case, the mobility tensor can be reduced to a scalar, and Eq. (1) can be written as:

$$\mu = \mu_{xx} = \mu_{yy} = \frac{\mu_{xx} + \mu_{yy}}{2} = \frac{q}{n_c \Omega} \sum_n \int \frac{d\mathbf{k}}{\Omega_{BZ}} \frac{\partial f_{n\mathbf{k}}}{\partial E_{n\mathbf{k}}} \tau_{n\mathbf{k}} \frac{v_{n\mathbf{k},x}^2 + v_{n\mathbf{k},y}^2}{2} = \frac{q}{n_c \Omega} \sum_n \int \frac{d\mathbf{k}}{\Omega_{BZ}} \frac{\partial f_{n\mathbf{k}}}{\partial E_{n\mathbf{k}}} \tau_{n\mathbf{k}} \frac{v_{n\mathbf{k}}^2}{2} \tag{3}$$

Since we only consider phonon-induced scattering in this paper, the $\tau_{n\mathbf{k}}$ can be calculated as:

$$\frac{1}{\tau_{n\mathbf{k}}} = \frac{2\pi}{\hbar} \sum_{mv} \int_{BZ} \frac{d\mathbf{q}}{\Omega_{BZ}} |g_{mnv}(\mathbf{k},\mathbf{q})|^2 [(f_{m\mathbf{k}+\mathbf{q}} + n_{v\mathbf{q}}) \delta(E_{n\mathbf{k}} - E_{m\mathbf{k}+\mathbf{q}} + \hbar\omega_{v\mathbf{q}}) \\ + (1 + n_{v\mathbf{q}} - f_{m\mathbf{k}+\mathbf{q}}) \delta(E_{n\mathbf{k}} - E_{m\mathbf{k}+\mathbf{q}} - \hbar\omega_{v\mathbf{q}})](1 - \cos\theta_{\mathbf{k},\mathbf{k}+\mathbf{q}}), \tag{4}$$

where the initial electronic state $n\mathbf{k}$ is scattered to the final state $m\mathbf{k}+\mathbf{q}$ by interacting with a phonon $v\mathbf{q}$ with frequency $\omega_{v\mathbf{q}}$ ($v$ is the phonon band index and $\mathbf{q}$ is the phonon wavevector); $n$ is the Bose distribution; $g_{mnv}(\mathbf{k},\mathbf{q})$ is the electron-phonon coupling (EPC) matrix element, defined as:

$$g_{mnv}(\mathbf{k},\mathbf{q}) = \langle \psi_{m\mathbf{k}+\mathbf{q}} | V_{v\mathbf{q}} | \psi_{n\mathbf{k}} \rangle. \tag{5}$$

The $\psi_{n\mathbf{k}}, \psi_{m\mathbf{k}+\mathbf{q}}$ are initial electronic state and final state, respectively. The $V_{v\mathbf{q}}$ is perturbation potential induced by phonon $v\mathbf{q}$, which can be calculated using density functional perturbation theory (DFPT).

As mentioned in the introduction section, in order to get the converged mobility, one needs to calculate a large amount of EPC matrix elements with a rather dense grid to sample the Brillouin zone (hundreds of points along each dimension), which is computationally very expensive if fully using DFPT. A more common approach is to interpolate the matrix elements from a sparse and DFPT-calculated grid to a dense grid. In principle, due to the nonanalyticity of perturbation potential in insulators and semiconductors when $\mathbf{q} \to 0$, the Wannier interpolation fails to replicate the EPC matrix elements for long-wavelength phonons. A widely used solution for this problem is to separate the short-range (SR) and long-range (LR) contributions to the perturbation potential [28–32]:

$$V_{\nu\mathbf{q}} = V_{\nu\mathbf{q}}^{SR} + V_{\nu\mathbf{q}}^{LR} \tag{6}$$

and thus the EPC matrix element becomes:

$$\begin{aligned}g_{mn\nu}(\mathbf{k},\mathbf{q}) &= \langle\psi_{m\mathbf{k}+\mathbf{q}}|V_{\nu\mathbf{q}}^{SR}|\psi_{n\mathbf{k}}\rangle + \langle\psi_{m\mathbf{k}+\mathbf{q}}|V_{\nu\mathbf{q}}^{LR}|\psi_{n\mathbf{k}}\rangle \\ &= g_{mn\nu}^{SR}(\mathbf{k},\mathbf{q}) + g_{mn\nu}^{LR}(\mathbf{k},\mathbf{q})\end{aligned} \tag{7}$$

The separation should satisfy that $V^{SR}$ is smooth and analytic so that the $g^{SR}$ can be Wannier interpolated well to a fine grid. The non-analytic component in $V$ is therefore fully contained in $V^{LR}$. In practice, one derives an explicit expression for the $V^{LR}$ and their $g^{LR}$, and subtracts it from the DFPT-calculated $g$ to get the $g^{SR}$. The $g^{SR}$ is then Wannier interpolated to the fine grid points, and the corresponding $g^{LR}$ is calculated from the derived expression. Finally, the $g^{LR}$ and $g^{SR}$ are summed up to recover the full $g$ on the fine grid.

The key to this process is the expression for $V^{LR}$ or $g^{LR}$. For 3D crystals, the leading-order of $g^{LR}$ due to the dipole (i.e. Fröhlich) perturbation potential is solved in Ref. [28,37], and the next-leading-order part (i.e. quadrupole) is addressed recently by several groups [29–32]. For 2D crystals, however, there are several issues: (1) First, when performing DFPT calculation with 3D periodic boundary conditions to get the $V$ and $g$ on a coarse grid, there are fictitious interactions between periodic images along the direction perpendicular to the basal plane of 2D material. This leads to some physical results, for example, the divergency of the $g$ for longitudinal optical (LO) phonon at $\mathbf{q}\rightarrow 0$, even if there is a large vacuum space between periodic images. To solve this problem, one needs to use the "Coulomb cutoff" technique, which truncates the fictitious interactions and makes the $g_{LO}$ converge at $\mathbf{q}\rightarrow 0$. [33,38] (2) The electrical screening is significantly different between 2D and 3D systems. Therefore, the $g^{LR}$ expressions for 3D cannot be directly applied to 2D. The dipolar $g^{LR}$ in 2D has been worked out by Ref. [33], which is particularly important for the LO mode at $\mathbf{q}\rightarrow 0$. However, the expression for quadrupole $g^{LR}$ is still missing, and its effect on the interpolation and mobility needs to be studied.

Recently, Royo and Stengel [36] developed a theory of the long-range electrostatic interactions in 2D crystals. Using interatomic force constants as examples, they showed that their method gives a better interpolation of the phonon band structure. Below, we will follow their method, and make one step forward to derive an explicit first-principles expression for the $g^{LR}$ that includes both the dipolar and quadrupole scatterings.

Ref. [36] shows that if the Coulomb kernel $v$ is separated into a LR part and a SR part (so that the singularity in $v$ is extracted to the LR component):

$$v = v_{\nu\mathbf{q}}^{LR} + v_{\nu\mathbf{q}}^{SR} \tag{8}$$

then the screened perturbation potential will be:

$$V^{SR} = W^{SR}\rho^{ext}$$
$$V^{LR} = (\varepsilon^{SR})^{-1}W^{LR}\rho^{SR}.$$
(9)

Here $\rho^{ext}$ is the "external" charge perturbation due to the nuclear displacement, which generates a "dressed" charge perturbation $\rho^{SR}$ through the short-range dielectric function $\varepsilon^{SR}$:

$$\rho^{SR} = [(\varepsilon^{SR})^{-1}]^{\dagger}\rho^{ext}.$$
(10)

$W$ is the screened Coulomb interaction:

$$W^{SR} = (\varepsilon^{SR})^{-1}v^{SR}$$
$$W^{LR} = (\varepsilon^{LR})^{-1}v^{LR}.$$
(11)

Further, it is shown that both $(\varepsilon^{SR})^{-1}$ and $(\varepsilon^{LR})^{-1}$ are related with an intermediate polarizability function $\chi^{SR}$:

$$(\varepsilon^{SR})^{-1} = (1+v^{SR}\chi^{SR})$$
$$(\varepsilon^{LR})^{-1} = (1-v^{LR}\chi^{SR})^{-1}.$$
(12)

Therefore, in principle, once we specify $v^{LR}$ and know $\chi^{SR}$, we can use Equations (9), (11) and (12) to obtain $V^{LR}$. Formally, the $\chi^{SR}$ can be calculated from irreducible polarizability $\chi^0$ by $\chi^{SR} = \chi^0(1-v^{SR}\chi^0)^{-1}$ but in practice it is more convenient to use an approximate form [36].

With the quantities provided by Ref. [36], neglecting the "electric field term" in $V^{LR}$, and using the phonon-induced displacement, we obtain an explicit expression for $V^{LR}$ (see SM Appendix A for details). Then using Eqs. (7), we get the long-range EPC matrix element $g^{LR}$ for 2D crystal with in-plane mirror plane as:

$$g_{mn\nu}^{LR}(\mathbf{k},\mathbf{q}) = \frac{e^2}{2\Omega}\sum_{\kappa}\left(\frac{\hbar}{2M_{\kappa}\omega_{v\mathbf{q}}}\right)^{1/2} f(q)\left[\frac{i\mathbf{q}\cdot\mathbf{Z}_{\kappa}^{\parallel}(\mathbf{q})\cdot\mathbf{e}_{\kappa\nu}(\mathbf{q})}{q\varepsilon^{\parallel}(\mathbf{q})}\langle\psi_{m\mathbf{k}+\mathbf{q}}|e^{i\mathbf{q}\cdot(\mathbf{r}-\boldsymbol{\tau}_{\kappa})}\cosh(qz)|\psi_{n\mathbf{k}}\rangle \right.$$
$$\left. + \frac{\mathbf{Z}_{\kappa}^{\perp}(\mathbf{q})\cdot\mathbf{e}_{\kappa\nu}(\mathbf{q})}{\varepsilon^{\perp}(\mathbf{q})}\langle\psi_{m\mathbf{k}+\mathbf{q}}|e^{i\mathbf{q}\cdot(\mathbf{r}-\boldsymbol{\tau}_{\kappa})}\sinh(qz)|\psi_{n\mathbf{k}}\rangle\right],$$
(13)

where $\Omega$ is the unit cell area, $\kappa$ is the index of the atom in the unit cell, $M$ is the atomic mass, $\mathbf{e}_{\nu}(\mathbf{q})$ is the eigenvector of phonon $\nu\mathbf{q}$, $\boldsymbol{\tau}$ is the atomic position. The $\varepsilon^{\parallel}, \varepsilon^{\perp}$ are in-plane and out-of-plane components of $\varepsilon^{LR}$, respectively. The $f(q)=1-\tanh(qL/2)$ is range separation function where $L$ is range separation length (a parameter used to define the $v^{LR}$ (see the SM Appendix D1). The Cartesian coordinate $z$ is perpendicular to the basal plane of 2D crystal and $z=0$ corresponds to the center of the 2D crystal.

The $\mathbf{Z}_\kappa(\mathbf{q})$ is a key matrix describing charge response to nuclear displacement and can be expanded around $\mathbf{q}=0$ as:

$$Z_{\kappa\alpha}^{\parallel,\beta}(\mathbf{q}) = \hat{Z}_{\kappa\alpha}^{(\beta)} - i\sum_\gamma \frac{q_\gamma}{2}\left(\hat{Q}_{\kappa\alpha}^{\beta\gamma} - \delta_{\beta\gamma}\hat{Q}_{\kappa\alpha}^{zz}\right)$$
$$Z_{\kappa\alpha}^{\perp}(\mathbf{q}) = \hat{Z}_{\kappa\alpha}^{(z)} - i\sum_\beta q_\beta \hat{Q}_{\kappa\alpha}^{z\beta}. \qquad (14)$$

Here $\alpha$, $\beta$, $\gamma$ are direction indices that go through $x$, $y$ and $z$ ($x$ and $y$ are parallel to the basal plane of the 2D crystal, while $z$ is perpendicular), and $\hat{Z}$ and $\hat{Q}$ are Born charges and quadrupole responses for 2D system (note that they have to be re-constructed from the standard output of the supercell calculation, see Appendix B in Ref. [36] for details). If $\hat{Q}$ in Equation (14) is set to zero, then the $g^{\text{LR}}$ will only contain dipolar scattering.

In practice, Eq. (13) can be simplified to a more compact form. First, the $\langle\psi_{m\mathbf{k}+\mathbf{q}}|e^{i\mathbf{q}\cdot(\mathbf{r}-\boldsymbol{\tau}_\kappa)}\sinh(qz)|\psi_{n\mathbf{k}}\rangle$ term in Equation (13) is negligible due to the in-plane mirror symmetry of the 2D crystal and the odd function of $\sinh(qz)$. Secondly, the 2D crystal can be assumed to be thin enough so the $\cosh(qz)\sim 1$ in Eq. (13). Therefore, the $g^{\text{LR}}$ can be approximated by:

$$g_{mn\nu}^{\text{LR}} = \frac{ie^2}{2\Omega}\sum_\kappa \left(\frac{\hbar}{2M_\kappa \omega_{\nu\mathbf{q}}}\right)^{1/2} f(q)\frac{\mathbf{q}\cdot\mathbf{Z}_\kappa(\mathbf{q})\cdot\mathbf{e}_{\kappa\nu}(\mathbf{q})}{q\varepsilon^{\parallel}(\mathbf{q})}\langle\psi_{m\mathbf{k}+\mathbf{q}}|e^{i\mathbf{q}\cdot(\mathbf{r}-\boldsymbol{\tau}_\kappa)}|\psi_{n\mathbf{k}}\rangle, \qquad (15)$$

In practice, we perform first-principles calculations using the Quantum Espresso Package [39,40] with SG15 Optimized Norm-Conserving Vanderbilt (ONCV) pseudopotentials [41,42] and the Perdew-Burke-Ernzerhof (PBE) exchange-correlation functional [43]. The kinetic energy cutoffs for $MoS_2$ and InSe are set to 70 and 90 Ry, respectively. The density functional perturbation theory (DFPT) calculations [25] are performed on $12\times 12$ $\mathbf{q}$ sparse grids with Coulomb cutoff technique [38]. We modified the EPW package [44,45] so that the $g^{\text{LR}}$ is calculated using Equation (15) and the $g^{\text{SR}}$ is Wannier interpolated from $12\times 12$ sparse $\mathbf{k}/\mathbf{q}$ grids to $300\times 300$ fine $\mathbf{k}/\mathbf{q}$ grids for $MoS_2$ ($600\times 600$ for InSe). To obtain the quadrupoles ($\hat{Q}$), we use DFPT to compute the macroscopic charge responses to different atomic displacements at several $\mathbf{q}$ points around $\mathbf{q}=0$, and perform a second order polynomial fitting (see Appendix C in SM for details). For the intrinsic semiconductor/insulator without free carriers, the $\varepsilon^{\parallel}$ can be well approximated as [36]:

$$\varepsilon^{\parallel}(\mathbf{q}) = 1 + 2\pi q \alpha^{\parallel}(\mathbf{q}) f(q), \tag{16}$$

where $\alpha^{\parallel}(\mathbf{q})$ is in-plane polarizability (which depends on direction $\mathbf{q}$ for anisotropic material). To obtain the intrinsic mobility, we place the Fermi level at the middle of the band gap. For doped semiconductor, it can be proved that the free carriers will mainly change the $V^{LR}$ by modifying $\varepsilon^{\parallel}$ (which can be obtained from QEH model [46,47]), while leaving the $V^{SR}$ largely intact (see Appendix D for details). Therefore, to study the doped semiconductor, one does not need to re-perform the coarse-grid DFPT calculation for the material with explicit additional electrons/holes. Instead, one can just use the DFPT-calculated $g$ for the intrinsic material; while after the Wannier interpolation of $g^{SR}$, the $g^{LR}$ of doped material instead of that of intrinsic material is added to $g^{SR}$ to obtain the total $g$ on fine grids. This method significantly reduces the computational cost compared to the approach of performing DFPT calculations for each carrier concentration.

**Results**:

First we evaluate the role of the quadrupoles in the interpolation of the EPC. We consider the electronic states located at an iso-energy circle of 30 meV above conduction band minimum (CBM), as they have a significant contribution to the mobility. We select one of the states as the initial state, and the others as the final. Figure 1b-f compare the interpolated and the DFPT-calculated EPC "strength" $D_{mn\nu}(\mathbf{q}) = \sqrt{\omega_{\mathbf{q}\nu}} |g_{mn\nu}(\mathbf{q})|$ [29,30] for intrinsic MoS$_2$ and InSe (here we use $D$ instead of $g$ to remove the common numerical instability in DFPT calculation for phonon frequency). For both materials, the initial state is selected to be 30 meV above the CBM along the Γ-M direction, and the final states are located at the iso-energy circle of the same valley. As can be seen in Figure 1, even with Coulomb cutoff and 2D dipoles, the interpolated $D$ (dashed lines) still significantly deviate from the DFPT calculated ones (dots), especially for TA mode for MoS$_2$ (Figure 1b), LA and TA modes for InSe (Figure 1d), and ZO$_1$ and ZO$_2$ modes for InSe (Figure 1e). After the incorporation of 2D quadrupoles, the interpolated $D$ (solid lines) become much closer to the DFPT calculated ones, indicating the importance of quadrupole scattering in EPC by those phonon modes. To assess the impact on the mobility, we calculate the intrinsic electron mobility at room temperature with and without quadrupoles. As shown in Figure 1a, for MoS$_2$, the mobility decreases from 167 to 136 cm$^2$V$^{-1}$s$^{-1}$ (by ~ 24%) when the quadrupoles are considered, while for InSe, the mobility increases from 103 to 117 cm$^2$V$^{-1}$s$^{-1}$, suggesting the non-negligible role of quadrupole scattering in mobility .

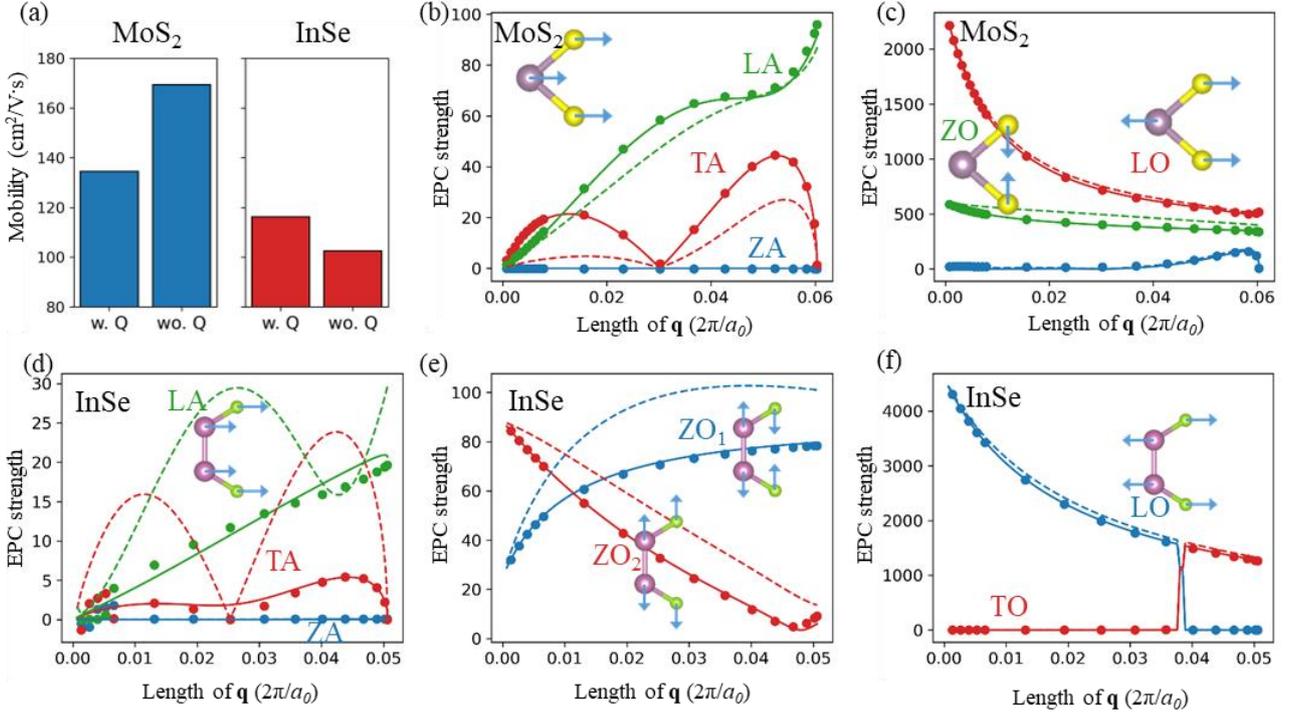

**Figure 1**. (a) Intrinsic electron mobility of MoS$_2$ and InSe, calculated with or without 2D quadrupoles. (b-f) comparison of the DFPT-calculated EPC strengths (dots) and the interpolated ones (dashed lines: without quadrupoles; solid lines: with quadrupoles), for selected phonon modes of intrinsic MoS$_2$ and InSe. For both materials, the initial state is selected to be 30 meV above the CBM along the Γ-M direction, and the final states are located at the iso-energy circle of the same valley. The vibration patterns are shown in the insets.

As an important extension to more realistic situation, we compute the EPC in doped 2D semiconductors with free electrons. As shown in Figure 2a, the increase in electron concentration ($n_e$) significantly changes the $D$ of TA and LO modes for MoS$_2$, and ZO$_1$, ZO$_2$ and LO modes for InSe. LO mode amongst has the most significant change (Figure 2c for InSe and Figure S2a for MoS$_2$): with free electrons, the $D$ of LO at long-wavelength limit reduces to zero, due to the screening of the Fröhlich potential in LO mode. If the mobility is limited by the Fröhlich EPC interaction, then it can change significantly with carrier concentration (as will be shown later). It is also interesting to find that, the $D$ of ZO$_2$ mode for InSe increases with $n_e$ (Figure 2b), which is apparently against the intuition that the carrier screening would weaken the EPC. This is because in this case, the $g^{LR}$ and $g^{SR}$ (both being the complex numbers) have opposite phases and the $g^{LR}$ has a smaller norm. As mentioned in the Methods section, the free carriers mainly screen the $g^{LR}$ while leave the $g^{SR}$ largely intact (in this work only the $g^{LR}$ is changed), thus the $|g| = |g^{LR}+g^{SR}|$ increases with $n_e$.

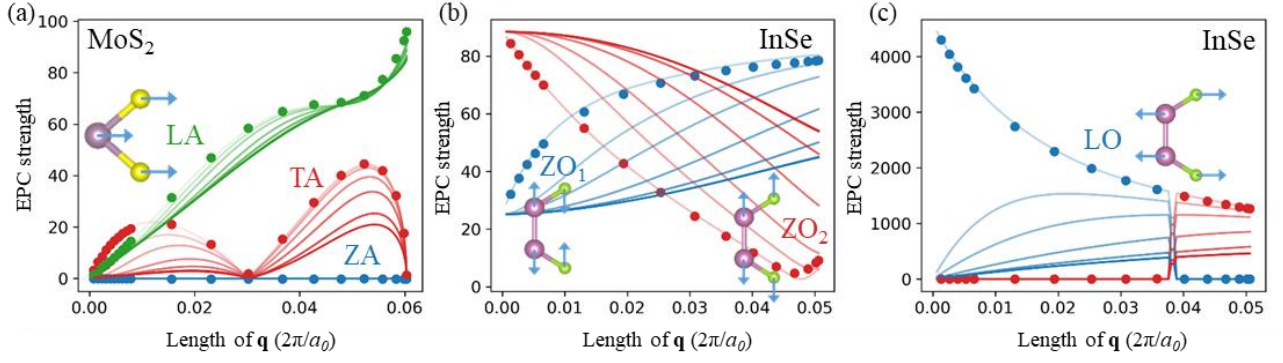

**Figure 2.** EPC strengths of $MoS_2$ acoustic phonon modes (a), InSe $ZO_1$ and $ZO_2$ modes (b), and InSe LO mode (c) over large carrier concentration range, from intrinsic to $3 \times 10^{13}$ cm$^{-2}$. The line thickness indicates the carrier concentration, with thicker lines for larger carrier concentration. The initial and final states are the same as those in Fig. 1.

With the carrier-screened EPC, we further compute the corresponding electron mobility in doped semiconductors. Figure 4a shows the calculated mobility of $MoS_2$ and InSe over a wide range of carrier concentration, from a low density of $n_e=10^{10}$ cm$^{-2}$, which corresponds to fermi energy (FE) at 140 meV below CBM for InSe and 182 meV below CBM for $MoS_2$, to a high density of $3 \times 10^{13}$ cm$^{-2}$, corresponding to FE at 320 meV above CBM for InSe and 72 meV above CBM for $MoS_2$. At $n_e=10^{10}$ cm$^{-2}$, the mobility is nearly identical to the intrinsic mobility (<3% differences for both InSe and $MoS_2$). With increasing carrier concentration, the mobility of $MoS_2$ does not change much (from 136 to the maximum of 172 cm$^2$V$^{-1}$s$^{-1}$), while that of InSe increases significantly, reaching 444 cm$^2$V$^{-1}$s$^{-1}$ at $10^{13}$ cm$^{-2}$ (3.8 times of the intrinsic mobility of 117 cm$^2$V$^{-1}$s$^{-1}$), and then decreases.

To understand why the InSe mobility is more sensitive to the carrier concentration, we calculated the LR mobility and SR mobility for both InSe and $MoS_2$. The LR mobility is calculated using $g^{LR}$ instead of total $g$ for Eq. 4, while the SR mobility is calculated from $g^{SR}$ only. As shown in Figure 3a, the $MoS_2$ has a lower SR mobility (196 cm$^2$V$^{-1}$s$^{-1}$) than its LR mobility (376 cm$^2$V$^{-1}$s$^{-1}$), while the InSe has an opposite order, with the LR mobility much lower than the SR mobility (120 vs 1099 cm$^2$V$^{-1}$s$^{-1}$). These suggest that the SR potential plays a more important role in limiting the $MoS_2$ mobility, while for InSe, the mobility is mainly limited by the LR potential. Indeed, as shown in Fig. 3b-c which compare the total scattering rates with the LR and SR scattering rates, the LR and SR potentials both contribute to the scattering in $MoS_2$, while for InSe, the scattering is dominated by the LR potential. Since LR potential is more significant in InSe than in $MoS_2$, and the free carriers mainly change the LR potential (see the methods section and the proof in Appendix D), InSe mobility should be more affected by free carriers. As a further demonstration, we show the scattering rates at $n_e=10^{13}$ cm$^{-2}$ in Fig. 3e-f. At such high concentration, the LR scattering is greatly suppressed for both $MoS_2$ and InSe, while the $MoS_2$ still has a significant SR scattering left, thus its mobility is lower than that of InSe.

The next question is why LR scattering is larger in InSe than that in MoS$_2$. Equation (14) and (15) indicate that the $g^{LR}$ is related to three material-dependent properties: Born charge ($\hat{Z}$), quadrupole response ($\hat{Q}$), and in-plane polarizability ($\alpha^\parallel$). The $\hat{Q}$ is less important than $\hat{Z}$ as it is in higher order of $q$ (Equation (14)). Our calculations show that InSe has a larger Born effective charge ($|\hat{Z}_{xx}|=|\hat{Z}_{yy}|\sim 2.5$ in InSe while 1 for Mo and 0.5 for S in MoS$_2$) and a smaller in-plane polarizability $\alpha^\parallel$ (4.8 Å in InSe and 7.0 Å in MoS$_2$), thus has a larger LR potential. As a further demonstration, in Figure 3d, we compare the |g| for the LO mode, which is the major source for the LR scattering in polar semiconductors, for intrinsic InSe and MoS$_2$. The choices of the initial and final states are the same as those for Fig. 1. Indeed, we find that the $|g_{LO}|$ in InSe is nearly 2.5 times larger than that of MoS$_2$. In addition to the difference in the magnitude of LR potential, another reason is the phonon wavelength involved in scattering. MoS$_2$ has two valleys (located at K and K') of the conduction band, while InSe has only one (located at Γ). Thus, MoS$_2$ has both intra-valley and inter-valley scatterings, which involve both long-wavelength and short-wavelength phonons, while InSe has only intra-valley scattering that involves the long-wavelength phonons only. Even for the intra-valley scattering, the shortest phonon wavelength in InSe is longer than that in MoS$_2$ (see Fig. 3d), due to the smaller band-edge effective mass of InSe. Therefore, the scattering in InSe involves more long-wavelength phonons than that in MoS$_2$. Since the LR potential is generated by the long-wavelength phonon, the LR scattering is hence more significant in InSe.

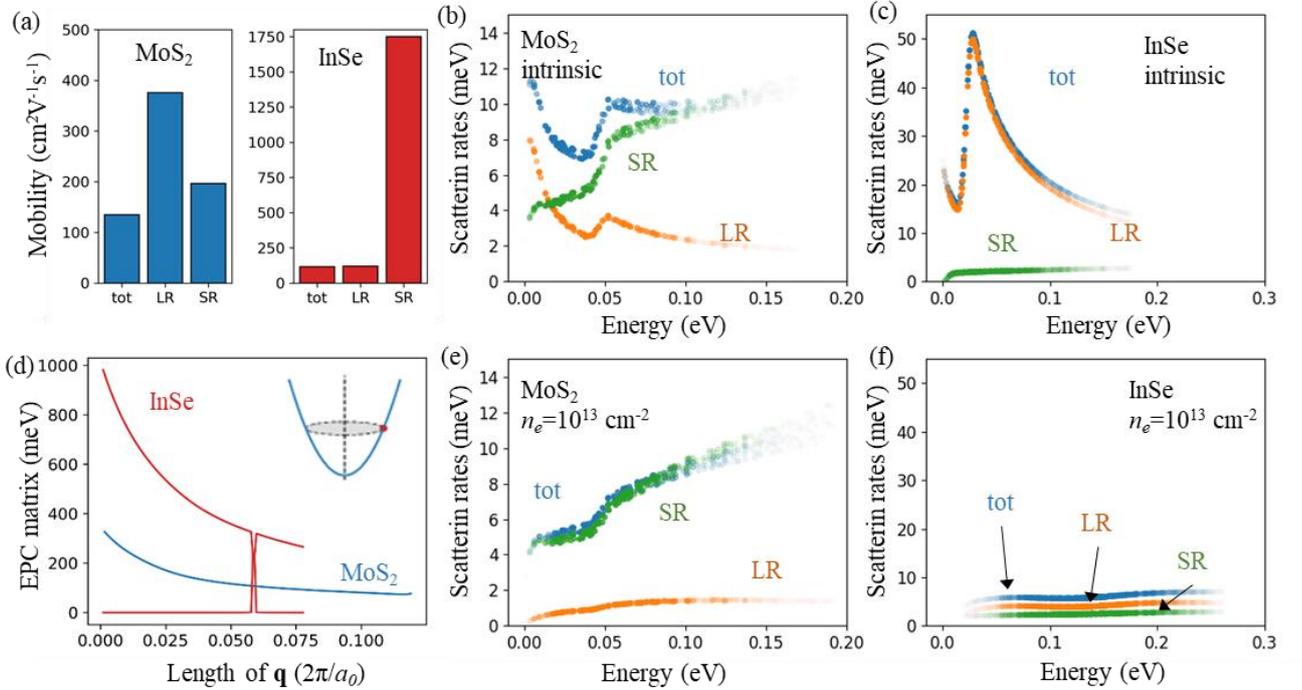

**Figure 3** (a) Electron mobility calculated using the total EPC matrix (tot), the long-range EPC only (LR), and the short-range only (SR) (see main text for definitions). (b), (c) Scattering rates for intrinsic MoS$_2$ and InSe. (d) The total EPC matrix element for LO mode of intrinsic MoS$_2$ and InSe.

The initial and final states are the same as those in Fig. 1, and are illustrated in the inset as well. (e), (f) Same as (b), (c) but for high carrier concentration ($n_e=10^{13}$ cm$^{-2}$).

Since the mobility of InSe shows a strong and non-monotonical change with carrier concentration, we focus on it and try to understand the origin of the non-monotonical change. Following our previous work [35,48], for 2D crystal with hexagonal lattice, we re-write Equation (3) in a form similar to Drude model:

$$\mu = |q|\bar{\tau}\langle M^{-1}\rangle, \quad (17)$$

where

$$\langle M^{-1}\rangle = \frac{1}{2n_e\Omega}\sum_n \int \frac{d\mathbf{k}}{\Omega_{BZ}} \frac{\partial f_{n\mathbf{k}}}{\partial E_{n\mathbf{k}}} v_{n\mathbf{k}}^2$$

$$\bar{\tau} = \frac{\mu}{|q|\langle M^{-1}\rangle} \quad . \quad (18)$$

The $\langle M^{-1}\rangle$ reflects the information of electronic structure, including the group velocities $v_{n\mathbf{k}}$ of the electronic states and their occupations $f_{n\mathbf{k}}$. The EPC information is wrapped in $\bar{\tau}$ by design. We call $\langle M^{-1}\rangle$ as "Drude inverse effective mass", and $\bar{\tau}$ as "Drude relaxation time".

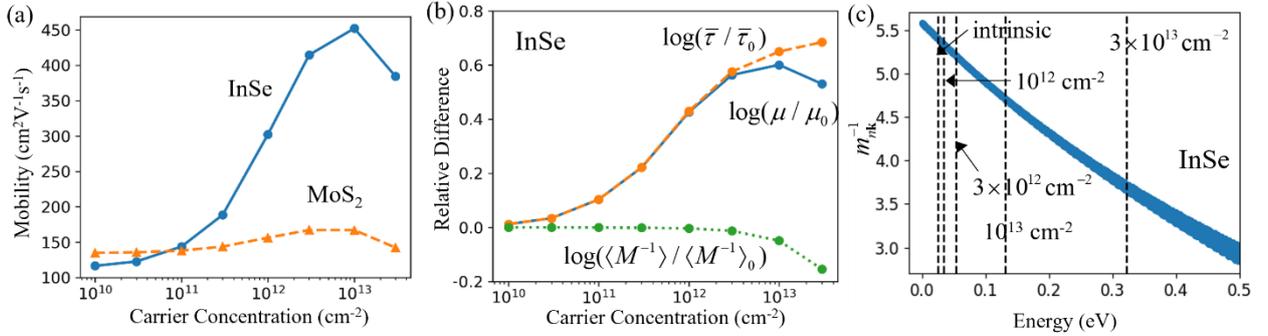

**Figure 4** (a) Electron mobility of InSe and MoS$_2$ over large carrier concentration range. (b) Logarithm change of the mobility ($\log(\mu/\mu_0)$), "Drude inverse effective mass" ($\log(\langle M^{-1}\rangle/\langle M^{-1}\rangle_0)$) and "Drude relaxation time" ($\log(\bar{\tau}/\bar{\tau}_0)$) with carrier concentration for InSe. (c) Local inverse effective mass $m_{n\mathbf{k}}^{-1}$ (see main text for definition) for InSe. The dashed lines indicate the energies where the electronic states are the most important to the mobility under different carrier concentrations.

The changes in $\bar{\tau}$, $\langle M^{-1}\rangle$ and $\mu$ of InSe vs. $n_e$ are shown is Figure 4b, where they are plotted in logarithm scale and referenced to the values for intrinsic InSe. It can be seen that for $n_e < 3\times 10^{12}$ cm$^{-2}$, the increase of $\mu$ is driven by the increase of $\bar{\tau}$, while the change in $\langle M^{-1}\rangle$ is negligible. For $n_e > 3\times 10^{12}$ cm$^{-2}$, although $\bar{\tau}$ still increases, the $\langle M^{-1}\rangle$ decreases, resulting in the decrease of $\mu$. The increase of $\bar{\tau}$ is consistent with the decrease of the scattering rates in Figure 3c and f, due to the free carrier screening. To understand the origin of the decrease of the $\langle M^{-1}\rangle$, in the Appendix E we prove that:

$$\langle M^{-1}\rangle = \frac{1}{n_e\Omega}\sum_n\int\frac{d\mathbf{k}}{\Omega_{BZ}}f_{n\mathbf{k}}\frac{1}{2}\left(\frac{\partial^2 E_{n\mathbf{k}}}{\hbar^2\partial k_x^2}+\frac{\partial^2 E_{n\mathbf{k}}}{\hbar^2\partial k_y^2}\right), \tag{19}$$

where $\frac{1}{2}\left(\frac{\partial^2 E_{n\mathbf{k}}}{\hbar^2\partial k_x^2}+\frac{\partial^2 E_{n\mathbf{k}}}{\hbar^2\partial k_y^2}\right) = m_{n\mathbf{k}}^{-1}$ can be interpreted as the "local" inverse effective mass for electronic state $n\mathbf{k}$. When weighted by $f_{n\mathbf{k}}/(n_e\Omega)$, the average of the local inverse effective mass over all the electronic states gives the Drude inverse effective mass. If the band structure is fully parabolic, then $m_{n\mathbf{k}}^{-1}$ should be a constant independent on the state. However, in reality, the $m_{n\mathbf{k}}^{-1}$ often decreases as the energy increases, as shown in Fig. 4c. In other words, the electrons become heavier at higher energies. When the $n_e$ increases, the $f_{n\mathbf{k}}/(n_e\Omega)$ weight term will make the higher energy states with lower $m_{n\mathbf{k}}^{-1}$ more occupied and contribute more to the $\langle M^{-1}\rangle$ (Fig. 4c), thus leading to decrease of $\langle M^{-1}\rangle$. Overall, a higher carrier concentration screens the scattering while involves heavier electrons, resulting in a non-monotonical change of the mobility.

**Conclusion:**

In this work we presented an efficient and accurate first-principles method to calculate electron-phonon coupling for 2D materials. By incorporating the quadrupole scattering, a much better interpolation quality of EPC matrix and thus a more accurate mobility are achieved as exemplified by MoS$_2$ and InSe systems. This method also allows for a natural incorporation of the effects of the free carriers, enabling a facile calculation of the screened EPC and the mobility for doped semiconductors. Our work reveals that the electron mobility of InSe is more sensitive to the free carriers than that of the MoS$_2$, due to its larger ratio of the long-range scattering that can be screened. As the electron concentration increases, the electron mobility of InSe can reach 444 cm$^2$V$^{-1}$s$^{-1}$, 3.8 times of the intrinsic mobility. However, further increase of the electron concentration decreases the mobility due to the involvement of heavier electronic states. We anticipate that the carrier-induced large improvement of the mobility can also be observed in other semiconductors whose scattering is dominated by the long-range scattering, such as polar ones with a single valley and/or small band-edge effective mass.

**Acknowledgements:**

Y. L. acknowledges the support of Welch Foundation (F-1959-20210327). The calculations were performed on the TACC cluster.

# Supplementary Material for "Phonon-limited Mobility of 2D Semiconductors: Quadrupole Scattering and Free-carrier Screening"

Chenmu Zhang, Yuanyue Liu*

*Texas Materials Institute and Department of Mechanical Engineering*

*The University of Texas at Austin, Austin, Texas 78712, USA*

Yuanyue.liu@austin.utexas.edu

**Appendix A:**

**A.1 Long-range EPC matrix $g^{LR}$**

From Equation (8)-(12) in the main text, the $V^{LR}$ can be obtained once the long-range Coulomb kernel $v^{LR}$ and short-range polarizability $\chi^{SR}$ are defined. Following Equation (20) in Ref. [1], the $V^{LR}$ due to a collective displacement of sublattice $\kappa$ along $\alpha$ direction $\Delta\tau_{\kappa\alpha}(\mathbf{R})=e^{i\mathbf{q}\cdot\mathbf{R}}$ can be expanded at long-wavelength limit:

$$V^{LR}_{\kappa\alpha}(\mathbf{r}) = \sum_{l,m=0,1} \varphi^{SR}_l(\mathbf{r}) \tilde{W}^{LR}_{lm} \tilde{\rho}^{SR}_{\kappa\alpha,m}, \tag{S1}$$

where $\varphi^{SR}$ is:

$$\begin{aligned}\varphi^{SR}_0(\mathbf{r}) &= \langle \mathbf{r} | (\varepsilon^{SR})^{-1} | \cosh(qz) \rangle \\ \varphi^{SR}_1(\mathbf{r}) &= \langle \mathbf{r} | (\varepsilon^{SR})^{-1} | \sinh(qz) \rangle\end{aligned} \tag{S2}$$

and $\tilde{W}^{LR}, \tilde{\rho}^{SR}_{\kappa\alpha}$ are:

$$\tilde{W}^{LR} = \frac{2\pi f(q)}{q} \begin{bmatrix} \frac{1}{1+2\pi f(q)\alpha^{\|}} & 0 \\ 0 & \frac{1}{1-2\pi f(q)\alpha^{\perp}} \end{bmatrix}; \quad \tilde{\rho}^{SR}_{\kappa\alpha} = \begin{bmatrix} -\frac{ie^{-i\mathbf{q}\cdot\boldsymbol{\tau}_\kappa}}{\Omega} \sum_\beta q_\beta Z^{\|,\beta}_{\kappa\alpha} \\ \frac{e^{-i\mathbf{q}\cdot\boldsymbol{\tau}_\kappa}}{\Omega} \sum_\beta q_\beta Z^{\perp,\beta}_{\kappa\alpha} \end{bmatrix}. \tag{S3}$$

The $f(q)$ is the range separation function $f(q)=1-\tanh(qL/2)$, $L$ is the range separation length, which should be larger than half of the 2D layer thickness. The definition of $Z$ can be found in Equation (14) in the main text.

The first approximation is to let $(\varepsilon^{SR})^{-1}=1+v^{SR}\chi^{SR}\sim 1$, which will lead to $\varphi^{SR}_0 \sim \cosh(qz), \varphi^{SR}_1 \sim \sinh(qz)$. The second term $v^{SR}\chi^{SR}$ corresponds to the the "electric-field term" in Ref. [2] and has negligible effect on mobility (0.1% for GaAs and 0.01% for GaP [2]). The "electric field term" describes the local-field potential induced by macroscopic electric field [2,3] and is usually disregarded in interpolation of phonon perturbation potentials [4] and $g$ matrix elements [5,6].

In order to consider the $V^{LR}$ induced by atom displacement corresponding to $v$-th phonon mode with wave vector **q**, we rewrite the $\Delta\tau$ as:

$$\Delta\boldsymbol{\tau}_{\kappa(\nu)} = \left(\frac{\hbar}{2M_\kappa \omega_{\nu\mathbf{q}}}\right)^{1/2} e^{i\mathbf{q}\cdot\mathbf{R}} \mathbf{e}_{\kappa\nu}(\mathbf{q}), \tag{S4}$$

where $\mathbf{e}_\nu(\mathbf{q})$ is the eigenvector of phonon $\nu\mathbf{q}$, $M_\kappa$ is the mass of atom $\kappa$ and $\omega$ is the phonon frequency. The corresponding $V^{LR}$ is:

$$V^{LR}_{\kappa(\nu)}(\mathbf{q},z) = \frac{2\pi f(q)}{q}\frac{\cosh(qz)}{1+2\pi q\alpha^{\parallel}(\mathbf{q})f(q)}\rho^{SR,\parallel}_{\kappa(\nu)}(\mathbf{q}) - \frac{2\pi f(q)}{q}\frac{\sinh(qz)}{1-2\pi q\alpha^{\perp}f(q)}\rho^{SR,\perp}_{\kappa(\nu)}(\mathbf{q}) \tag{S5}$$

and $\rho^{SR}$ is:

$$\rho^{SR,\parallel}_{\kappa(\nu)}(\mathbf{q}) = \frac{-ie^{-i\mathbf{q}\cdot\tau_\kappa}}{\Omega}\left(\frac{\hbar}{2M_\kappa\omega_{\nu\mathbf{q}}}\right)^{1/2} \mathbf{q}\cdot\mathbf{Z}^{\parallel}_\kappa(\mathbf{q})\cdot\mathbf{e}_{\kappa\nu}(\mathbf{q}),$$

$$\rho^{SR,\perp}_{\kappa(\nu)}(\mathbf{q}) = \frac{e^{-i\mathbf{q}\cdot\tau_\kappa}}{\Omega}\left(\frac{\hbar}{2M_\kappa\omega_{\nu\mathbf{q}}}\right)^{1/2} q\mathbf{Z}^{\perp}_\kappa(\mathbf{q})\cdot\mathbf{e}_{\kappa\nu}(\mathbf{q}). \tag{S6}$$

The total $V^{LR}$ corresponding to phonon $\nu\mathbf{q}$ is:

$$V^{LR}_{\nu\mathbf{q}} = \sum_\kappa V^{LR}_{\kappa(\nu)}. \tag{S7}$$

Therefore, from Equation (S5)-(S7) and definition of $g^{LR}$ (Eq. (7) in the main text), the $g^{LR}$ can be written as:

$$g^{LR}_{mn\nu}(\mathbf{k},\mathbf{q}) = \frac{e^2}{2\Omega}\sum_\kappa \left(\frac{\hbar}{2M_\kappa\omega_{\nu\mathbf{q}}}\right)^{1/2} f(q)\left[\frac{i\mathbf{q}\cdot\mathbf{Z}^{\parallel}_\kappa(\mathbf{q})\cdot\mathbf{e}_{\kappa\nu}(\mathbf{q})}{q[1+2\pi\alpha^{\parallel}(\mathbf{q})f(q)]}\langle\psi_{m\mathbf{k}+\mathbf{q}}|e^{i\mathbf{q}\cdot(\mathbf{r}-\tau_\kappa)}\cosh(qz)|\psi_{n\mathbf{k}}\rangle\right.$$
$$\left.+\frac{\mathbf{Z}^{\perp}_\kappa(\mathbf{q})\cdot\mathbf{e}_{\kappa\nu}(\mathbf{q})}{1-2\pi\alpha^{\perp}f(q)}\langle\psi_{m\mathbf{k}+\mathbf{q}}|e^{i\mathbf{q}\cdot(\mathbf{r}-\tau_\kappa)}\sinh(qz)|\psi_{n\mathbf{k}}\rangle\right], \tag{S8}$$

which is exact Equation (13) in the main text.

**A.2 Selection of range separation length $L$**

As shown in Equation (S8), the range separation function $f(q)$ enters the expression of $g^{LR}$. The parameter $L$ in $f(q)$ determines the boundary of $v^{LR}$ and $g^{LR}$. A smaller $L$ means more Coulomb potential in $v$ is divided into $v^{LR}$ and it will lead to a larger spread of $g^{LR}$ in reciprocal space. While a larger $L$ will lead to more Coulomb potential being considered as short-ranged and a larger percentage of $g$ being classified as $g^{SR}$ which will be Wannier interpolated from real space.

Therefore, an appropriate $L$ needs to be carefully selected. Based on our tests, a relatively small $L$ will give $g$ closer to DFPT benchmarks. In Ref. [1], the minimum of $L$ with $\varepsilon^\perp > 0$ is:

$$L_{min} = 4\pi\alpha^\perp. \tag{S9}$$

So, in the calculations, the $L$ is selected as:

$$L = 1.1 L_{min} = 4.4\pi\alpha^\perp. \tag{S10}$$

However, the $L$ selected by Equation (S10) will overestimate $g^{LR}$ when $\alpha^\perp$ is too small for certain materials. As shown in Figure S1, the $L$ given by Equation (S10) gives overestimated ZA $g$ matrix for Sb (See blue dashed line in Figure S1c), which is mainly due to the large $g^{LR}$ tails on $\mathbf{G}_\| \neq 0$ grids. The $f(q)$ is supposed to fast decay with increasing $|\mathbf{q}+\mathbf{G}_\||$ in reciprocal space and lead to vanishing $g^{LR}$ on $\mathbf{G}_\| \neq 0$ grids. Otherwise, the charge response $\mathbf{Z}$ and $\chi^{SR}$ need to be expended to higher order for large $\mathbf{q}+\mathbf{G}_\|$. So in practical calculations, a Gaussian function is used to better control the spread of $g^{LR}$ in reciprocal space: we use $\bar{g}^{LR} = \exp(-q^2/(2R))g^{LR}$ instead of $g^{LR}$ calculated from Equation (S8) in Wannier interpolation. The usage of Gaussian function in $\bar{g}^{LR}$ leads to nearly identical interpolated EPC strength $D$ in MoS$_2$ and InSe, as shown in Figure S1a, S1b. Moreover, using $\bar{g}^{LR}$ will lead to reasonable interpolated $D$ in Sb while the $g^{LR}$ gives fictitious $D$ for ZA mode, as shown in Figure S1c.

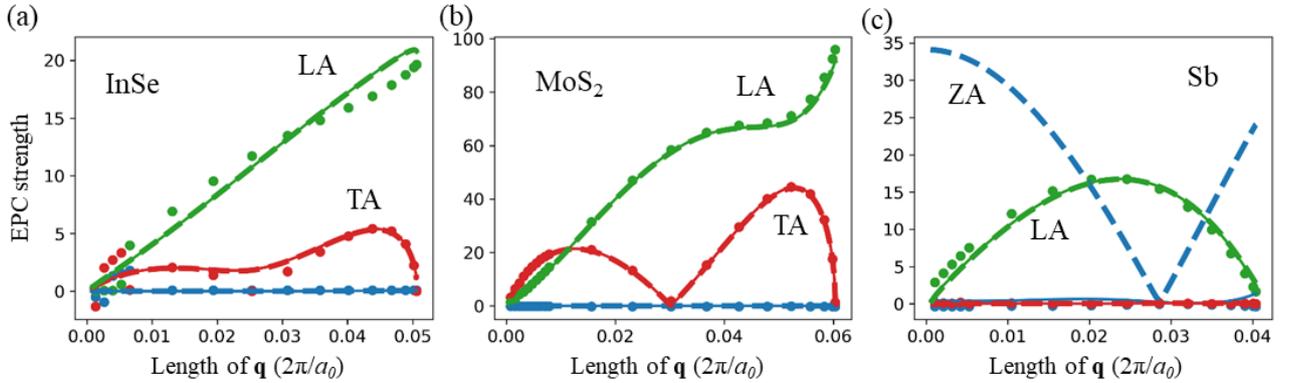

**Figure S1** Interpolated EPC strength using $\bar{g}^{LR}$ (a Gaussian function is multiplied based on $g^{LR}$) and $g^{LR}$ (defined in Equation (15) in the main text). Solid lines indicate EPC strength interpolated via $\bar{g}^{LR}$, dashed lines correspond to $g^{LR}$, and scatters are benchmarks from direct DFPT calculation for acoustic modes of InSe (a), MoS$_2$ (b) and Sb (c).

**Appendix B:**

**EPC strength with free carrier screening**

In Figure 2 in the main text, the interpolated EPC strengths with free carrier screening effect for acoustic modes in $MoS_2$ and optical modes for InSe are plotted. Here the rest of EPC strengths, optical modes for $MoS_2$ and acoustic modes for InSe, are supplemented in Figure S2.

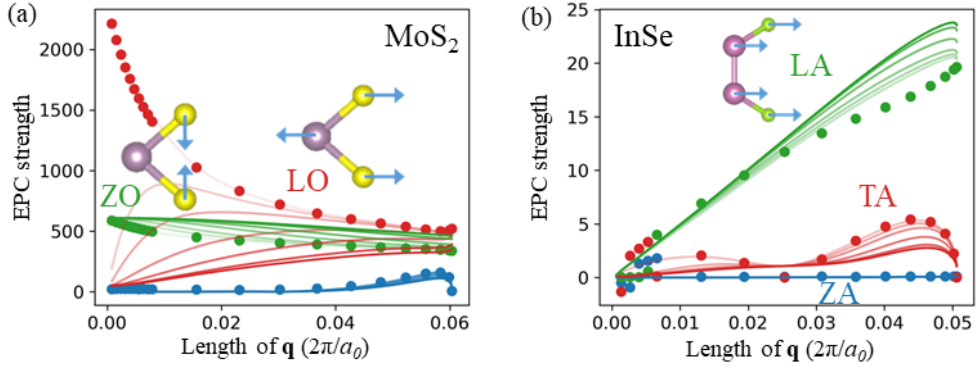

**Figure S2** The interpolated EPC strength of $MoS_2$ optical phonon modes (a), InSe acoustic modes (b) over large carrier concentration range, intrinsic to $3 \times 10^{13}$ cm$^{-2}$. The thickness of color indicates the carrier concentration, with thicker lines for larger carrier concentration.

**Appendix C:**

**Quadrupole from QE**

The key quantity, dynamical quadrupoles, are calculated via Quantum Espresso, in order to keep computational details consistent with EPC matrix elements. In principle, we followed the quadrupole calculation approach in literatures [7,8], extracting the electron charge density response from DFPT calculations on different **q** points and computing the quadrupoles by polynomial fitting. Recently, there is an algorithm implemented in ABINIT [9,10] based on current-density response, enabling the calculation for not only quadrupoles also the flexoelectric tensor. However, it brings complexity for coding and as far as we know, it is limited to LDA and not implemented in software other than ABINIT. Here we compare our results with the quadrupoles from literatures for bulk systems, Si, GaN and 2D materials, BN and $Sn_2S_2$, as an important verification of quadrupole tensors used in computations of $g^{LR}$. For bulk system, we used fitting equations from Equation (146)-(148) in Ref. [7]. For 2D systems, we used fitting equations from Appendix B of Ref. [1].

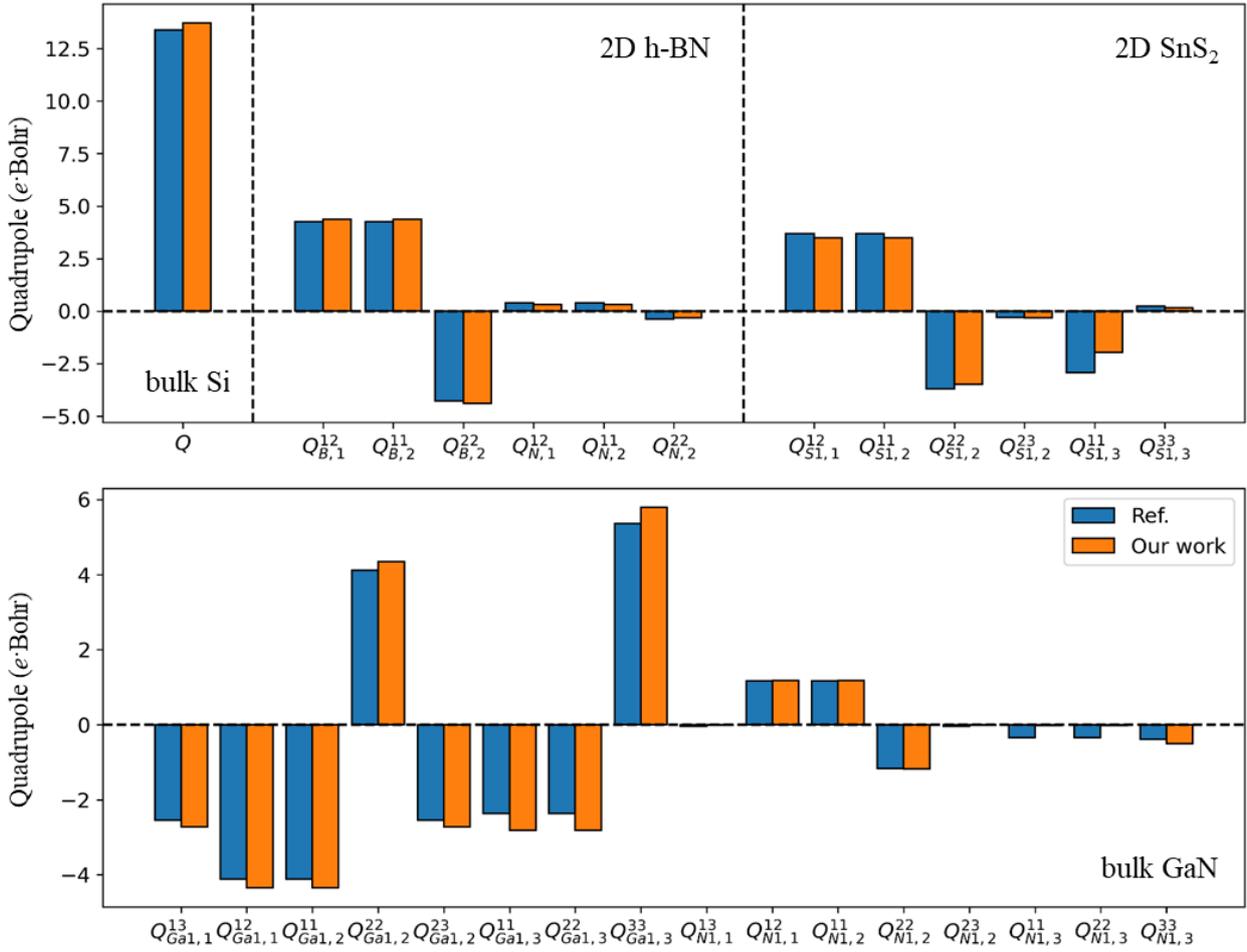

**Figure S3**. Quadrupole comparison between the calculated results of this work and literatures for Si [11], GaN [5], BN monolayer and SnS$_2$ monolayer [1]. Only nonzero elements in the quadrupole tensor are presented.

In above calculations, we perform first-principles calculations using the Quantum Espresso Package [12,13] with SG15 Optimized Norm-Conserving Vanderbilt (ONCV) pseudopotentials [14,15] and the Perdew-Burke-Ernzerhof (PBE) exchange-correlation functional [16]. The difference of quadrupoles for each tensor element is lower than 0.4 $e\cdot$Bohr, which might come from inevitable octupoles or higher order and relatively larger numerical errors in our method.

**Appendix D:**

**D.1 Coulomb kernel separation**

In Eq. (8) in the main text, the bare Coulomb kernel $v$ is separated into a short-range (SR) $v^{SR}$ and a long-range (LR) part $v^{LR}$. A desirable separation needs to divide the non-analyticity of $v$ into $v^{LR}$ and ensure the $v^{SR}$ is an analytic function of wave vector **q** in reciprocal space and decays fast in real space. In 3D systems, a proper definition of $v^{LR}$ is:

$$v^{LR}_{GG'}(\mathbf{q}) = \delta_{G0}\delta_{G'0} \frac{4\pi}{q^2}, \tag{S11}$$

which leaves the $v^{SR}$ as:

$$v^{SR}_{GG'}(\mathbf{q}) = (\delta_{GG'} - \delta_{G0}\delta_{G'0}) \frac{4\pi}{|\mathbf{q}+\mathbf{G}|^2}. \tag{S12}$$

The $v^{SR}$ is an analytic function of **q** in reciprocal space for different **G**, which brings many benefits and will be discussed in the following. However, for 2D systems, an appropriate Coulomb separation leading to an analytic $v^{SR}$ is not easy to find. In Ref. [Stengel], a proper separation regime is proposed as:

$$v^{LR}(\mathbf{K}_\parallel, z-z') = \frac{2\pi}{|K_\parallel|} f(|K_\parallel|) \cosh[K_\parallel(z-z')], \tag{S13}$$

where $\mathbf{K}_\parallel = \mathbf{q}+\mathbf{G}_\parallel$, **q** is the in-plane wave vector, $\mathbf{G}_\parallel$ is the in-plane reciprocal space Bravais lattice and $f(K_\parallel)$ is the range separation function defined as:

$$f(K_\parallel) = 1 - \tanh(\frac{K_\parallel L}{2}) \tag{S14}$$

and $L$ is range separation length. The $v^{LR}$ defined in Eq. (S13) will leave the remaining $v^{SR}$ analytic for 2D electrostatic interaction. The analyticity of $v^{SR}$ can be verified in reciprocal space with supercell context. Here we follow the Appendix C of Ref. [1] and display some key results which will be useful in later discussion. Suppose we have 2D crystals placed on plane $z=2nL$ ($n$ is integer) under 3D periodic boundary condition. $L$ is large enough to avoid interlayer charge density overlap. To avoid the interlayer electrostatic interaction, the Coulomb cutoff technique is required and the $v$ can be written as [17]:

$$v(\mathbf{r}) = \frac{\theta(L-|z|)}{r}, \tag{S15}$$

where **r** is the real space coordinates. The $v$ in reciprocal space is:

$$v(\mathbf{K}_\parallel, G_n) = \frac{4\pi}{K_\parallel^2 + G_n^2}[1-(-1)^n e^{-K_\parallel L}], \tag{S16}$$

where $G_n = n\pi/L$ is the Bravais lattice of $z$ direction. As shown in Appendix C in Ref. [1], the $v^{LR}$ in reciprocal space can be obtained from Fourier transform:

$$v^{LR}(\mathbf{K}_\parallel, G_n) = \int_{-L}^{L} v^{LR}(\mathbf{K}_\parallel, z) e^{iG_n z} dz = \frac{4\pi}{K_\parallel^2 + G_n^2}(-1)^n(1-e^{-K_\parallel L}). \tag{S17}$$

Then the $v^{SR}$ is:

$$v^{SR}(\mathbf{K}_{\|}, G_n) = \frac{4\pi}{K_{\|}^2 + G_n^2}[1-(-1)^n]. \tag{S18}$$

We can see when $v^{SR}$ is analytic at $\mathbf{q}=0$ for any $\mathbf{G}_{\|}$ and $G_n$.

### D.2 Properties of perturbation potential $V^{LR}$ and $V^{SR}$

In this section we show that the LR and SR perturbation potential $V^{LR}$, $V^{SR}$ have desirable properties for practical calculations. According to Equation (8)-(12) in the main text, the $V^{SR}$ and $V^{LR}$ can be written as:

$$\begin{aligned} V^{SR} &= (\varepsilon^{SR})^{-1} v^{SR} v^{-1} U^{ext} \\ V^{LR} &= (\varepsilon^{SR})^{-1} W^{LR} [(\varepsilon^{SR})^{-1}]^? v^{-1} U^{ext} \end{aligned} \tag{S19}$$

where $U^{ext}$ is the bare perturbation potential, due to phonon displacement in this article. Note that the $V^{SR}$ is determined by $(\varepsilon^{SR})^{-1}$ and $v^{SR}$. From Equation (12) in the main text, the $(\varepsilon^{SR})^{-1}$ is defined as:

$$(\varepsilon^{SR})^{-1} = 1 + v^{SR} \chi^{SR} = 1 + v^{SR} \chi^0 (1 - v^{SR} \chi^0)^{-1}, \tag{S20}$$

where $\chi^0$ is the irreducible polarizability, which is analytic at $\mathbf{q}=0$ for both metals and insulators [18]. Therefore, from Equation (S19) and (S20), the analytic of $V^{SR}$ is totally determined by $v^{SR}$. Once the $v^{SR}$ is analytic at $\mathbf{q}=0$, the $V^{SR}$ is also analytic at $\mathbf{q}=0$. The analyticity of $V^{SR}$ in 3D systems using Coulomb separation as Equation (S11) and (S12) has been discussed in Ref. [3]. The Equation (S20) here can be extended to different systems and $v^{SR}$.

As a step further, now we consider the effect of the $\chi^0$ change on $V^{SR}$ and $V^{LR}$. Suppose there is a perturbation of the system leading to the $\chi^0$ change as:

$$\chi^0 \to \chi^0 + \Delta\chi^0, \tag{S21}$$

where $\Delta\chi^0$ describes the difference of $\chi^0$ between the new system and the original one. A typical example is free carrier doping. The $\chi^0$ of doped semiconductor has intra-band (largely due to the free carriers; denoted as $\Delta\chi^0$) and inter-band (close to $\chi^0$ of intrinsic system) components [19,20]. Here we impose additional limitation on $\Delta\chi^0$:

$$v^{SR}\Delta\chi^0 = 0. \tag{S22}$$

From Equation (S20), it can be seen the change of $\chi^0$ in Equation (S21) will not change the $(\varepsilon^{SR})^{-1}$ and thus the $V^{SR}$. This means if the change of screening properties can be approximated by $\Delta\chi^0$ satisfying Equation (S22), then the $V^{SR}$ will remain the same and the change of screened perturbation potential $V$ will be limited to $V^{LR}$, which will be convenient since the $V^{LR}$ has explicit approximated equation (Equation (S5)).

**D.3 Free carrier screening approximation in 3D and 2D**

In this section we discuss the exact free carrier screening model satisfying the Equation (S22), and the consequently change in $V^{LR}$ for both 3D and 2D systems. For 3D case, the free carrier screening effect can be approximated by $\Delta\chi^0$ with:

$$\Delta\chi^0 = \delta_{\mathbf{G}0}\delta_{\mathbf{G'}0}\chi^{3DEG}(\mathbf{q}), \tag{S23}$$

where $\chi^{3DEG}$ is the 3D free electron gas polarizability which can be calculated from Lindhard dielectric function. The Equation (S23) means the change of screening properties in doped semiconductor is considered only via adding an additional term $\chi^{3DEG}$ on the head term of dielectric matrix $\varepsilon$ of intrinsic material. From Equation (S12) and (S23), it can be seen the $v^{SR} \Delta\chi^0 = 0$. As discussed above, therefore the $(\varepsilon^{SR})^{-1}$ and thus the $V^{SR}$ will not change under the free carrier screening approximation in Equation (S23).

Then we consider the exact change of $V^{LR}$ due to $\Delta\chi^0$. From Equation (S19), the only change of $V^{LR}$ exists in $W^{LR}$:

$$W^{LR} = (1 + v^{LR}\chi)v^{LR}. \tag{S24}$$

From definition of $v^{LR}$ in Equation (S11), the $W^{LR}$ is:

$$W^{LR}_{\mathbf{GG'}}(\mathbf{q}) = \delta_{\mathbf{G}0}\delta_{\mathbf{G'}0}\frac{4\pi}{q^2}[1 + \frac{4\pi}{q^2}\chi_{00}] = \delta_{\mathbf{G}0}\delta_{\mathbf{G'}0}\frac{4\pi}{q^2}\varepsilon^{-1}_{00}(\mathbf{q}) = \delta_{\mathbf{G}0}\delta_{\mathbf{G'}0}\frac{4\pi}{q^2\epsilon_M(\mathbf{q})}, \tag{S25}$$

where $\varepsilon_M(\mathbf{q})$ is the macroscopic dielectric function. Therefore, the $V^{LR}$ of doped semiconductors can be determined once we know the corresponding $\varepsilon_M(\mathbf{q})$.

For 2D system, the free carrier screening effect can be approximated by $\Delta\chi^0$:

$$\Delta\chi^0 = \delta_{n0}\delta_{n'0}\chi^{2DEG}_{\mathbf{G}_\parallel \mathbf{G}'_\parallel}(\mathbf{q}), \tag{S26}$$

where $\chi^{2DEG}$ is the irreducible polarizability of 2D free electron gas. From definition in Equation (S18), the $v^{SR}=0$ when $n=0$. Therefore, $v^{SR}\Delta\chi^0 = 0$ and thus the $(\varepsilon^{SR})^{-1}$ and $V^{SR}$ will not change under the change: $\chi^0 \rightarrow \chi^0 + \Delta\chi^0$, which is similar to the 3D case.

Then we compute the $W^{LR}$ in 2D, from definition in Equation (S13), the $v^{LR}$ can be written as:

$$v^{LR}(\mathbf{K}_\|, z-z') = \begin{bmatrix} \cosh(K_\| z) & \sinh(K_\| z) \end{bmatrix} \begin{bmatrix} \dfrac{2\pi f(K_\|)}{K_\|} & 0 \\ 0 & -\dfrac{2\pi f(K_\|)}{K_\|} \end{bmatrix} \begin{bmatrix} \cosh(K_\| z') \\ \sinh(K_\| z') \end{bmatrix} \quad (S27)$$

or in matrix form:

$$v^{LR}(\mathbf{K}_\|, z-z') = \boldsymbol{\varphi}(z) \cdot \tilde{v}^{LR}(K_\|) \cdot \boldsymbol{\varphi}(z'), \quad (S28)$$

where $\boldsymbol{\varphi}(z) = [\cosh(K_\| z), \sinh(K_\| z)]$ and $\tilde{v}^{LR}(K_\|)$ is the $2\times 2$ matrix in Equation (S27). The $W^{LR}$ can be written in a similar form:

$$W^{LR}(\mathbf{q}, z-z') = \boldsymbol{\varphi}(z) \cdot [\tilde{v}^{LR}(q) + \tilde{v}^{LR}(q) \cdot \tilde{\chi}(\mathbf{q}) \cdot \tilde{v}^{LR}(q)] \cdot \boldsymbol{\varphi}(z'), \quad (S29)$$

where the $\mathbf{G}_\| \neq 0$ components are neglected due to the $f(K_\|) \sim 0$ when $\mathbf{G}_\| \neq 0$. The 4 components of $\tilde{\chi}$ can be computed as:

$$\begin{aligned}
\tilde{\chi}_{00} &= \int_{-L}^{L} dz \int_{-L}^{L} dz' \cosh(qz)\cosh(qz') \chi_{00}(\mathbf{q}, z, z') \\
\tilde{\chi}_{01} &= \int_{-L}^{L} dz \int_{-L}^{L} dz' \sinh(qz)\cosh(qz') \chi_{00}(\mathbf{q}, z, z') \\
\tilde{\chi}_{10} &= \int_{-L}^{L} dz \int_{-L}^{L} dz' \cosh(qz)\sinh(qz') \chi_{00}(\mathbf{q}, z, z') \\
\tilde{\chi}_{11} &= \int_{-L}^{L} dz \int_{-L}^{L} dz' \sinh(qz)\sinh(qz') \chi_{00}(\mathbf{q}, z, z')
\end{aligned} \quad (S30)$$

The key quantities are reducible polarizability elements $\tilde{\chi}_{\alpha\beta}, \alpha, \beta \in \{0,1\}$ which can be obtained from QEH codes. In QEH codes, the corresponding quantities $\hat{\chi}_{i\alpha}$ is computed as [21]:

$$\hat{\chi}_{i\alpha}(q,\omega) = \iint dz\, dz' (z-z_i)^\alpha \chi_i(z,z',q,\omega)(z-z_i)^\alpha, \quad (S31)$$

where $i$ is the layer index, $z_i$ is position of center of the 2D crystal and $\chi_i$ is the $\mathbf{G}_\|=\mathbf{G}'_\|=0$ component of reducible polarizability of layer $i$. Note that there are several slight differences between the $\tilde{\chi}_{\alpha\beta}$ defined in Equation (S30) and $\hat{\chi}_{i\alpha}$ in QEH model (Eq. (S31)). First, the off-diagonal term of $\hat{\chi}_{i\alpha}$ is neglected since most 2D materials have in-plane symmetry. Then the $\cosh(qz)$ in Equation (S30) is approximated as 1 and $\sinh(qz)$ is computed by $(z-z_i)$ in Eq. (S31) approximately. The latter will lead to a coefficient difference of $q^2$: $\tilde{\chi}_{11} \approx q^2 \hat{\chi}_1$, since $\sinh(qz) \sim qz$.

Now that the $W^{LR}$ can be computed using $\hat{\chi}_{i\alpha}$, which is calculated from first-principles via QEH codes [21,22]. Following the same procedure in Appendix A, the corresponding $\varepsilon^{\|}$ is:

$$\frac{1}{\varepsilon^{\|}} = 1 + \frac{2\pi}{q} f(q)\hat{\chi}_0. \tag{S32}$$

Combining Equation (S8) and (S32), the $g^{LR}$ and mobility $\mu$ for intrinsic materials can be computed based on $\hat{\chi}_{i\alpha}$ from QEH codes. Note that the $\varepsilon^{\|}$ shown above is different from the Equation (16) in the main text, where $\alpha^{\|}(\mathbf{q})$ is a constant extracted from supercell dielectric constant. The intrinsic mobilities calculated from $\varepsilon^{\|}$ using Equation (S32) and Equation (16) give very close $g$ matrices and $\mu$ (2% $\mu$ difference in $MoS_2$ and 3% in InSe), verifying the feasibility of dealing with dielectric screening in $g^{LR}$ via QEH model. In QEH model, the free carrier screening effect can be incorporated in $\hat{\chi}_{i\alpha}$ by including contributions from intra-band transitions caused by the free carriers. The free carrier contributions modify the $\chi^0$ only on monopole component (i.e. $n=n'=0$), which meets the requirement $v^{SR} \Delta\chi^0 = 0$.

**D.4 Justification of free carrier screening approximation**

Above we showed that the $V^{SR}$ and $V^{LR}$ in doped semiconductor can be conveniently computed ($V^{SR}$ remains the same as in intrinsic materials, $V^{LR}$ only changes proportionally) once the free carrier screening effect follows $v^{SR} \Delta\chi^0 = 0$, which usually means the free-carrier effect on $\varepsilon_{\mathbf{GG'}}$ confines on $\mathbf{G}=\mathbf{G'}=0$ head term (See Equation (S23) and (S26)). However, in realistic materials, the free carrier will also change the wings and body part of $\varepsilon_{\mathbf{GG'}}$ (See Equation (17) in Ref. [20]), which is assumed to be small and neglected in above derivation. The above "local-field" free-carrier screening approximation is also used in [19,22,23].

**Appendix E:**

**Proof of effective Drude equation**

Assuming the relaxation time is constant $\bar{\tau}$, the carrier mobility tensor in Equation (1) of the main text can be written as:

$$\begin{aligned}
\mu_{\alpha\beta} &= \frac{q\bar{\tau}_{\alpha\beta}}{n_e\Omega} \sum_n \int \frac{d\mathbf{k}}{\Omega_{BZ}} v_{n\mathbf{k},\alpha} v_{n\mathbf{k},\beta} \frac{\partial f_{n\mathbf{k}}}{\partial \varepsilon_{n\mathbf{k}}} \\
&= \frac{q\bar{\tau}_{\alpha\beta}}{n_e\Omega} \sum_n \int \frac{d\mathbf{k}}{\Omega_{BZ}} v_{n\mathbf{k},\alpha} \frac{\partial f_{n\mathbf{k}}}{\partial \hbar k_\beta} \\
&= -\frac{q\bar{\tau}_{\alpha\beta}}{n_e\Omega} \sum_n \int \frac{d\mathbf{k}}{\Omega_{BZ}} \frac{\partial v_{n\mathbf{k},\alpha}}{\partial \hbar k_\beta} f_{n\mathbf{k}} \\
&= -\frac{q\bar{\tau}_{\alpha\beta}}{n_e\Omega} \sum_n \int \frac{d\mathbf{k}}{\Omega_{BZ}} m^{-1}_{n\mathbf{k},\alpha\beta} f_{n\mathbf{k}} \\
&= q\bar{\tau}_{\alpha\beta} \langle M^{-1} \rangle_{\alpha\beta}.
\end{aligned} \quad (S33)$$

where $m^{-1}_{n\mathbf{k},\alpha\beta}$ is the inverse effective mass tensor. Here we reduce the tensors $\mu_{\alpha\beta}, \bar{\tau}_{\alpha\beta}, \langle M^{-1} \rangle_{\alpha\beta}$ to scalars $\mu$, $\bar{\tau}$ and $\langle M^{-1}\rangle$ since they share the same crystal symmetry of MoS$_2$ and InSe. Then for isotropic materials like MoS$_2$ and InSe we have:

$$\begin{aligned}
\mu &= \frac{1}{2}(\mu_{xx} + \mu_{yy}) = -\frac{q\bar{\tau}}{n_e\Omega} \sum_n \int \frac{d\mathbf{k}}{\Omega_{BZ}} \frac{1}{2}\left(m^{-1}_{n\mathbf{k},xx} + m^{-1}_{n\mathbf{k},yy}\right) f_{n\mathbf{k}} \\
&= -\frac{q\bar{\tau}}{n_e\Omega} \sum_n \int \frac{d\mathbf{k}}{\Omega_{BZ}} m^{-1}_{n\mathbf{k}} f_{n\mathbf{k}}.
\end{aligned} \quad (S34)$$

Then we obtain the Drude-like equation: $\mu = q\bar{\tau}\langle M^{-1}\rangle$, where $\langle \cdots \rangle$ denotes the average over Brillouin Zone weighted by fermi distribution $f$. Since $\langle M^{-1}\rangle$ is only determined by band structure and $f$ (i.e. carrier concentration), the $\bar{\tau}$ can be quantified by a pair of *ab initio* $\mu$ and $\langle M^{-1}\rangle$.

**Appendix F:**

**Analysis on eigenvectors of LR matrix elements**

In this section we provide more details on LR matrix elements $g^{LR}$ and their corresponding phonon vibration eigenvectors. The $g^{LR}$ can be separated into dipolar $g^{LR}$, only containing dipoles $\hat{Z}$ and quadrupolar $g^{LR}$, which is directly due to the quadrupoles $\hat{Q}$ in Equation (14) and (15) in the main text. First we compare the magnitude of EPC strengths ($D_{mn\nu}(\mathbf{q}) = \sqrt{\omega_{\mathbf{q}\nu}} |g_{mn\nu}(\mathbf{q})|$) of dipolar, quadrupolar and total LR EPC strengths $D^{LR}$ for acoustic (AC) modes in MoS$_2$ (Figure S4a) and InSe (Figure S4b). The $D^{LR}$ containing only quadrupoles (dotted lines), only dipoles (dashed lines) and both quadrupoles and dipoles (i.e. total LR part, as solid lines) are compared with DFPT benchmarks (dots).

The first thing worth noticing is that the dipolar $D^{LR}$ in LA/TA modes has the same order of $q$ as the quadrupolar $D$ near $\mathbf{q}=0$. This is due to the dipolar $D^{LR}$ in LA/TA originates from the LA/TA and LO phonon mixing [3]. Although the quadrupoles term in charge response $\mathbf{Z}$ (See Equation (14) for definition) is in order of $q^2$, while the dipole term is in order of $q$. The standard LO component in realistic LA/TA modes is proportional to $q$, leading to the same order of $q$ for both dipolar $D^{LR}$ and quadrupole $D^{LR}$ in LA/TA modes. In Figure S4d, the calculated LA and LO eigenvectors are projected on normalized (or standard) LO mode. An obvious linear standard LO component in the LA eigenvector exists in MoS$_2$ and confirms that the LA/LO mixing leads to the dipolar $D^{LR}$ in LA mode. The dipolar $D^{LR}$ is larger in InSe than MoS$_2$ due to the larger born effective charges (2.6 for InSe and 0.5 for MoS$_2$).

It is also interesting to compare the magnitude of different $D^{LR}$ and to find out how dipolar $D^{LR}$ and quadrupolar $D^{LR}$ contribute to the total $D^{LR}$. As shown in Figure S4a, the total $D^{LR}$ (solid lines) is larger than dipole (dashed lines) and quadrupole terms (dotted lines) in MoS$_2$, while in InSe (Figure S4b), the total $D^{LR}$ is much smaller than dipolar $D^{LR}$ or quadrupolar $D^{LR}$ due to their canceling out. Therefore, although both MoS$_2$ and InSe show relatively low $D^{LR}$ in LA mode, they are due to different reasons. In MoS$_2$, both the dipolar $D$ and the quadrupole $D$ are relatively weak compared to short-range EPC strength $D^{SR}$; while in InSe, the dipolar $D^{LR}$ and quadrupolar $D^{LR}$ give opposite macroscopic electric field and cancel each other, thus leading to relatively weak $D^{LR}$ compared to $D^{SR}$.

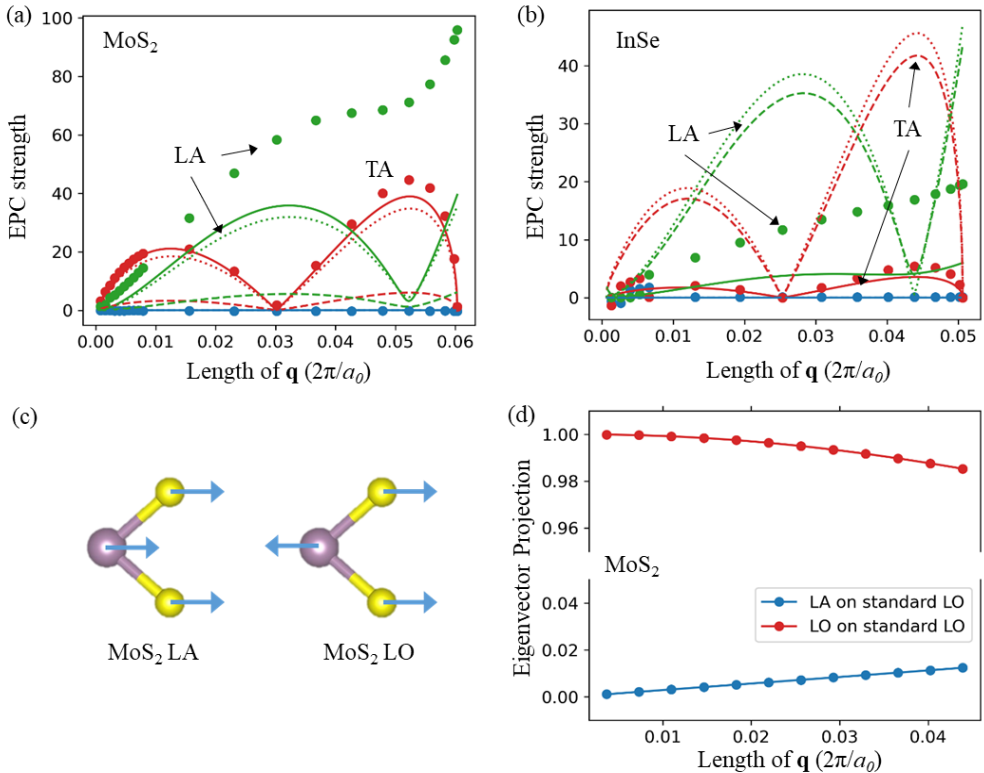

**Figure S4** (a) Interpolated long-range EPC strength containing only quadrupoles (dotted lines), only dipoles (dashed lines) and both quadrupoles and dipoles (solid lines) for MoS$_2$ acoustic modes. The scatters show total EPC strength calculated from DFPT calculations. (b) Same as (a) but for InSe. (c) Diagrams of standard LA and LO vibration eigenvectors for MoS$_2$. (d) LA and LO eigenvector projection on standard LO eigenvector for MoS$_2$.

**Appendix G**

**Coulomb cutoff influence on EPC strengths**

In the main text, we compared the $D$ calculated with 2D Coulomb cutoff and dipolar and quadrupolar $g^{LR}$ (i.e. "w. Q" in Figure 1 in the main text), and $D$ calculated with 2D Coulomb cutoff and dipolar $g^{LR}$ (i.e. "wo. Q" in Figure 1) in Figure 1. Here we compare the $D$ calculated with dipolar $g^{LR}$ but without 2D Coulomb cutoff (i.e. "3D wo. Q") with $D$ from "wo.Q" regime in Figure S5. A $12 \times 12$ **k/q** sparse grid is used in Wannier interpolation of $g^{SR}$. Comparing Figure 1 and Figure S5, it can be seen that the Coulomb cutoff is less important than quadrupoles for accurately computing $D$ in both MoS$_2$ and InSe. The incorporation of Coulomb cutoff improves the EPC interpolation of TO mode in MoS$_2$ (Figure S5c) and ZO$_2$ mode in InSe (Figure S5f). This might be due to the DFPT calculations in this article are performed on a $12 \times 12$ **q** grid, which is too sparse for Coulomb cutoff to make significant difference on perturbation potentials. Although the Coulomb cutoff seems not important for EPC interpolation in Figure S5, it will be necessary to ensure a converged EPC with a denser **k/q** grid.

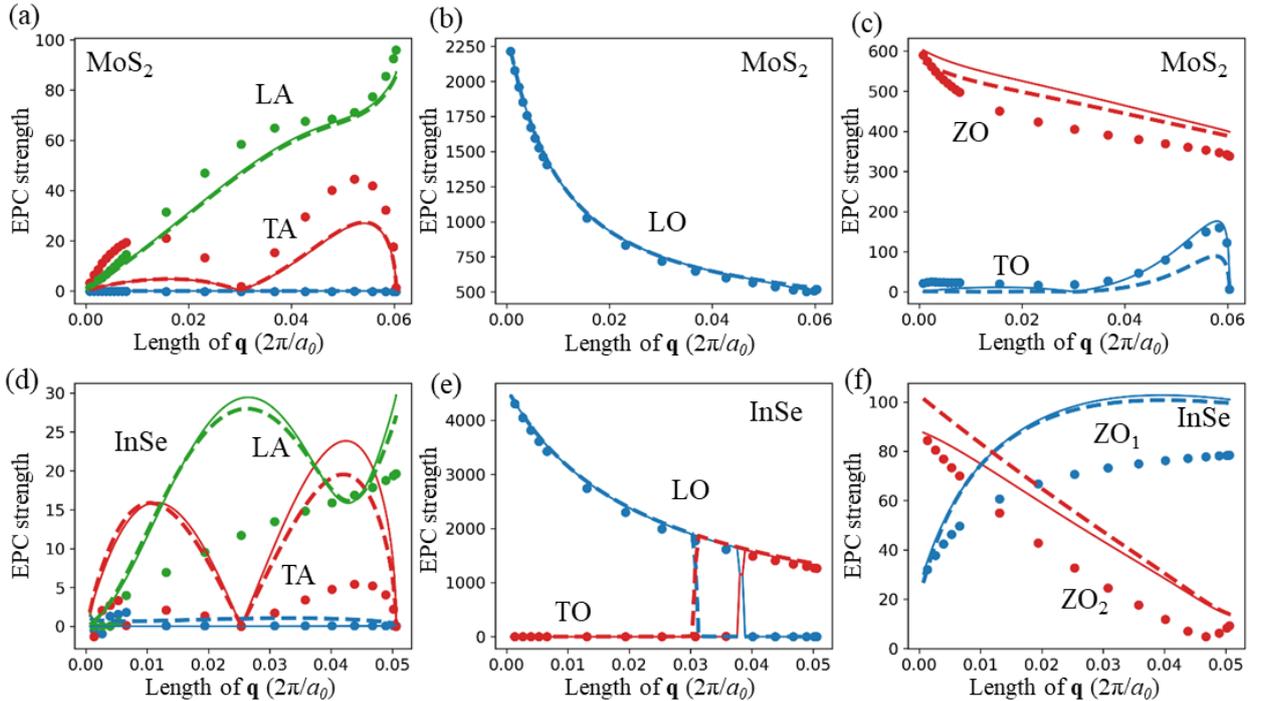

**Figure S5.** EPC strength comparison between different interpolation methods, with solid lines for Wannier interpolation with dipoles and 2D cutoff ("wo. Q"), dashed lines for Wannier interpolation with dipole and standard DFPT calculation for bulk system ("3D wo. Q"), and scatters for DFPT benchmarks, for MoS$_2$ (a), (b), (c) and InSe (d), (e), (f).